\documentclass[fleqn,twoside]{article}
\usepackage{espcrc2}
\usepackage{graphicx}

\title{The spectrum features of UHECRs below and surrounding GZK}

\author{V.P. Egorova, A.V. Glushkov, A.A. Ivanov\thanks{Presenting author.
E-mail: a.a.ivanov@ikfia.ysn.ru}, S.P. Knurenko, V.A. Kolosov, A.D. Krasilnikov,
I.T. Makarov, A.A. Mikhailov, V.V. Olzoev, M.I. Pravdin, A.V. Sabourov, I.Ye.
Sleptsov and G.G. Struchkov\address{Yu.G. Shafer Institute of Cosmophysical
Research and Aeronomy, Yakutsk 678980, Russia}}

\begin{document}

\begin{abstract}
The energy spectrum of ultra-high energy cosmic rays (UHECRs) is discussed on
the basis of the Yakutsk array database analysis. In the region $E_0=10^{17}$ to
$3\times10^{19}$ eV the showers are detected under trigger-500, while at
energies above $3\times 10^{19}$ eV the whole acceptance area for trigger-1000
is used in order to utilize all the data available in the region of
Greisen-Zatsepin-Kuzmin (GZK) cutoff. Updated relations are adapted to estimate
the primary energy having detected densities of charged particles at the
distance 300/600 m from the shower axis.
\vspace{1pc}
\end{abstract}

\maketitle

\section{Introduction}
The primary energy estimation method is one of the clue features to deal with
using extensive air shower (EAS) technique of ultra-high energy cosmic ray
measurements. In this paper we discuss the present status of algorithms used to
assign the primary energy to showers detected with the Yakutsk array. Special
attention is paid to the total flux measurement of the Cherenkov light on the
ground which is the basis of the model-independent approach to the energy
evaluation of EAS original particle. The uncertainty limits of the energy
estimation method are analyzed.

Resulting energy spectrum of cosmic rays in the region surrounding GZK energy
$E_{GZK}\sim 4\times10^{19}$ eV has the form consistent with the prediction of
Greisen \cite{G} and Zatsepin and Kuzmin \cite{ZK}. One more EAS event at energy
above $10^{20}$ eV has been detected with the Yakutsk array 18 February 2004. An
updated lateral distribution function (LDF) of charged particles was used in
data analysis which leads to increased $S_{600}$ parameter in average. As the
result we have 3 showers of energy greater than $10^{20}$ eV detected in Yakutsk
to present day.

\section{Data acquisition and analysis}
Actually, the Yakutsk array has two kinds of triggers to select the showers from
the background: produced by Cherenkov light detectors and by the scintillators.
We will describe here the latter one; in the case of coincident signal (particle
density $\rho>0.5\: m^{-2}$) in two scintillators of each station within 2
microseconds the signal passes on to the central controller. Trigger-500 is then
produced in the case of coincident signal (in 40 $\mu$s) from three or more
stations at $\sim500$ m spacing. Similarly, trigger-1000 is produced by $\sim 1$
kilometer spacing stations. After 1992 when 18 new stations were added, the
array area is increased from $2.5\: km^2$ to $7.2\: km^2$ where trigger-500
operates. That is why we can deal with the spectrum in the energy range from
$3\times10^{17}$ eV to $3\times10^{19}$ eV using the same trigger.

In our previous analysis the lateral distribution of charged particles was
approximated by Greisen's formula \cite{Cat1988}:
\begin{equation}
f(r)=(\frac{r}{r_0})^{-1.0}(1+\frac{r}{r_0})^{1.0-b},
\label{Eq:Greisen}
\end{equation}
where $b=1.38+2.16\cos\theta+0.15\log_{10}S_{600}$;

here we use another approximation given earlier \cite{Afan}:
\begin{equation}
f(r)=(\frac{r}{r_0})^{-1.3}(1+\frac{r}{r_0})^{1.3-b}(1+\frac{r}{2000})^{-c},
\label{Eq:Afan}
\end{equation}
where parameter $c$ has the value given in Table~\ref{table:index}, and $b$ is
additionally adjusted;

\noindent which has been shown to describe better the experimental points at
large core distances $r>1000$ m. As a result, the charged particle density at
600 m from the core, $S_{600}$, is increased by $\sim 10\%$ for showers above
$2\times 10^{19}$ eV with axes inside array perimeter, and by $\sim 20\%$
outside.

\begin{table}
\caption{Lateral distribution index}
\label{table:index}
\begin{tabular}{@{}ccccc}
\hline
$\log_{10}E_0$ & $<18.0$ & 18.1 & 18.3 & $>18.7$ \\
\hline
 $c$    &   0   &  1.6 &  2.3 &   3.5 \\
\hline
\end{tabular}\\[2pt]
\end{table}

\section{Energy estimation}
The main distinctive feature of the Yakutsk array is the air Cherenkov light
measurement. The total flux of the light emitted in atmosphere, $Q_{tot}$, is
used as the main estimator of the primary particle energy. In order to derive
the relation between $Q_{tot}$ and ionization loss of the shower electrons in
atmosphere we have to use the classic formula by Tamm and Frank \cite{Tamm} for
the number of Cherenkov photons in the wavelength interval
($\lambda_1,\lambda_2$) emitted when relativistic electron moves in air $l$ cm
with the speed $v$:
\begin{equation}
a(x,E)=2\pi\alpha l(\frac{1}{\lambda_2}-\frac{1}{\lambda_1})(1-\frac{c^2}{v^2n}),
\label{Eq:Tamm}
\end{equation}
where $\alpha=1/137$; $n$ is refraction coefficient. Emission is possible when
the particle velocity is greater than $c/n$. At the depth $x$ in atmosphere the
threshold energy of electrons is $E_{thr}=mc^2/\sqrt{2(n_0-1)\rho(x)/\rho_0}$,
where $n_0,\rho_0$ are air refraction coefficient and density at the sea level.
In the first approximation with regard to $mc^2/E$ the number of photons emitted
on the length $1\:g/cm^2$ is
\begin{equation}
a(x,E)\approx 4\pi\alpha(\frac{1}{\lambda_2}-\frac{1}{\lambda_1})\frac{n_0-1}{\rho_0}(1-\frac{E^2_{thr}}{E^2}).
\label{Eq:Tamm1}
\end{equation}
The total light flux at the sea level is given as the integral of electron energy spectrum in the shower $f(E_0,E,x)=-\partial N_e/\partial E$:
\begin{eqnarray}
Q_{tot}(E_0,x_0)=\nonumber\\
=-\int_0^{x_0}dx k(x,x_0)\int_{E_{thr}}^{E_0}dE f(E_0,E,x)a(x,E),
\end{eqnarray}
where $k(x,x_0)$ is attenuation coefficient of the light intensity in atmosphere. Integrating by parts and using the boundary conditions $a(x,E_{thr})=0,\:N_e(E_0,E_0,x)=0$ we take
\begin{eqnarray}
Q_{tot}(E_0,x_0)=\nonumber\\
=\int_0^{x_0}dx k(x,x_0)\int_{E_{thr}}^{E_0}dE N_e(E_0,E,x)a'(x,E).
\end{eqnarray}
Due to slow variable $N_e(E_0,E,x)$ at $E_{thr}$ and $a'(x,E)\propto E^2_{thr}/E^3$, we can  factor $N_e(E_0,E_{thr},x)$ outside the integral over energy:
\begin{eqnarray}
Q_{tot}(E_0,x_0)\approx4\pi\alpha(\frac{1}{\lambda_2}-\frac{1}{\lambda_1})\times\nonumber\\
\times\frac{n_0-1}{\rho_0}\int_0^{x_0}dx k(x,x_0)N_e(E_0,E_{thr},x).
\end{eqnarray}
The energy spectrum of electrons near the shower maximum gives $N_e(E_0,E_{thr},x)=N_e(E_0,0,x)\times\eta$, where $\eta=1-\epsilon e^\epsilon\int_\epsilon^\infty dt e^{-t}/t$ is independent of $x$; $\epsilon=2.29E_{thr}/\beta$. The cascade curve of electrons can be approximated according to the measurements by HiRes \cite{HiResCurve}:
\[
N_e(E_0,0,x)=N_{max}exp(-\frac{2}{\sigma^2}(\frac{x-x_{max}}{x+2x_{max}})),
\]
where $\sigma=0.272\pm0.002$. Attenuation of the Cherenkov light in atmosphere has the length $\Lambda \gg(x_0-x)$
\cite{Mono}. Substituting $exp((x_0-x)/\Lambda)$ to the total flux integral and expanding near $x_{max}$ we have
\begin{eqnarray}
Q_{tot}(E_0,x_0)\approx4\pi\alpha(\frac{1}{\lambda_2}-\frac{1}{\lambda_1})\frac{n_0-1}{\rho_0}\times\nonumber\\
\times\eta k(x_{max},x_0)(1+\delta Q)\int_0^{x_0}dx N_e(E_0,0,x),
\end{eqnarray}
$|\delta Q|\leq5\%$. On the other hand, $E_i=\beta/t_0\int_0^{x_0}dx N_e(E_0,0,x)$ is ionization integral. The final relation is
\begin{eqnarray}
Q_{tot}(E_0,x_0)\approx4\pi\alpha(\frac{1}{\lambda_2}-\frac{1}{\lambda_1})\times\nonumber\\
\times\frac{n_0-1}{\rho_0}\eta k(x_{max},x_0)\frac{t_0E_i}{\beta}
\label{Eq:relate}
\end{eqnarray}
with the accuracy $\sim5\%$. It is determined by $x_{max}$ and attenuation of light in atmosphere but is independent of hadronic interaction model assumptions.

Hereafter we analyze the shower parameters governing the energy fractions transferred to EAS components.

\subsection{Energy balance of EAS components}
The energy fractions of EAS primary particle transferred to the shower components can be described on the basis of hadron transport equations \cite{Mono,Erlykin}. If $E_k,\: (k=N, \pi, \mu\nu, e\gamma)$ is the energy transferred to nucleons, charged pions, muons+neutrinos, electrons+photons, then we are going to demonstrate in this subsection, with simple arguments, that a few cascade parameters determine the ratios between $E_k$ - the energy balance in the
shower. For instance, the transport equation for charged pions density $\pi (x,E)$ at depth $x$:
\begin{eqnarray}
\frac{\partial\pi(x,E)}{\partial
x}=-(\frac{1}{\lambda_\pi}+\frac{B_\pi}{x E})\pi(x,E)+\nonumber\\
+\frac{2}{3\lambda_\pi}\int_{E}^{E_0}\pi(x,U)w_{\pi\pi}(E,U)dU +\nonumber\\
+\frac{2}{3\lambda_N}\int_{E}^{E_0}N(x,U)w_{\pi N}(E,U)dU,
\end{eqnarray}
where interaction mean free paths $\lambda_{\pi},\lambda_N$ are
assumed constant; $w_{\pi\pi}(E,U),w_{\pi N}(E,U)$ are the spectra
of charged pions produced in pion-air and nucleon-air
interactions, can be transformed (integrating $\int EdE$) to:
\begin{equation}
\frac{dE_\pi}{dx}=-\frac{E_\pi}{\lambda_\pi}-\frac{B_\pi\pi(x,E>0)}{x}+
\frac{2E_\pi}{3\lambda_\pi}+\frac{2K_N E_N}{3\lambda_N},
\end{equation}
where $E_\pi(x)=\int_{0}^{E_0}\pi(x,E)EdE$;
$\pi(x,E>0)=\int_{0}^{E_0}\pi(x,E)dE$; $K_N$ is nucleon
inelasticity supposed to be constant; $E_N=E_0 exp(-K_N
x/\lambda_N)$.

In the energy range $E>>B_\pi$ the only parameters to define the quantity $E_\pi$ are $K_N/\lambda_N$ and $\lambda_\pi$. It means that the energy transferred to charged pions is independent of the spectra of pions produced in nuclear interactions. Hence, in the general case, we can use the simple $\delta$-model with $w_{ik}(E,U)=n_s\delta(E-U/n_s)$, where $n_s$ is the multiplicity of secondaries, to balance the components energy in a shower. Because of the net value of $E_{\mu+\nu}/E_0\leq 0.1$, the uncertainty due to simplified model should be of the second order of magnitude.

\begin{figure}[htb]
\includegraphics[width=\columnwidth]{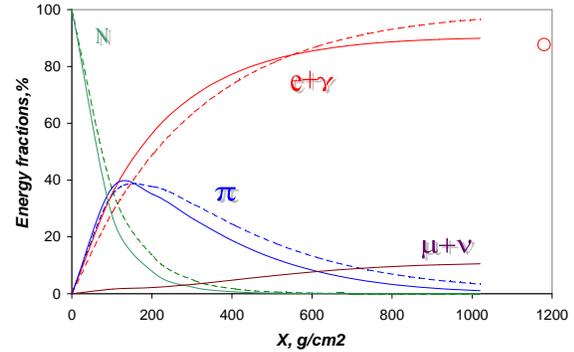}
\caption{The energy carried by the cascade nucleons ($N$), charged
pions ($\pi$), muons and neutrinos ($\mu+\nu$), electrons and
photons ($e+\gamma$). Dashed curves show analytic expressions with
$\lambda_i, K_N=const, B_\pi=0$: $E_N=E_0exp(-K_N x/\lambda_N)$;
$E_\pi=2/3E_0(1-exp(-K_Nx/\lambda_N))exp(-x/\lambda_\pi/3)$;
$E_{e\gamma}=E_0-E_N-E_\pi$. Solid curves are $\delta$-model
results: cross sections are supposed rising $\propto 0.08lnE$;
$B_\pi=120$ GeV; $K_N=0.5$; $n_s\propto E^{1/4}$; $E_0=10^{18}$
eV.}
\label{fig:balance}
\end{figure}

To summarize, the main model parameters to govern the energy balance in the shower are average inelasticity coefficients, mean free paths, multiplicity of secondaries and the fragmentation rate of primary nucleus. Other model characteristics such as 'the form of rapidity distribution of constituent quarks' are of less influence. The resultant energies in the case of constant $\lambda_{\pi},\lambda_N, K_N$ and $B_\pi=0$ are shown in Fig.~\ref{fig:balance} together with $\delta$-model results. An asymptotic ($x=\infty$) estimation of $E_{e\gamma}$ with CORSIKA(+QGSjet) program at $E_0=10^{18}$ eV (open circle) \cite{Song} is shown here as well.

\subsection{Experimental evaluation of the energy dispensed to EAS
components}
Energy fractions of the main EAS components can be estimated using the Yakutsk array data. Ionization loss of electrons is measured here detecting the total flux of the Cherenkov light on the ground. A relation between these values is given by formula (\ref{Eq:relate}) with parameters relevant to the Yakutsk array, taking into account detector calibration and atmospheric transparency factors \cite{Mono,CERN}:
\[
E_i=\frac{2.18\times10^4 Q_{tot}}{0.37+1.1\times10^{-3}X_{max}}.
\]

Other portions of the energy carried out by electromagnetic and muonic components beyond the sea level is evaluated via the total number of electrons:
\[
E_g=\epsilon_0 N_e\lambda_e/t_0,
\]
where $\epsilon_0, t_0$ are the critical energy and radiation length of electrons in air; attenuation length $\lambda_e$ is derived from zenith angle dependence of $N_e$ \cite{Afan};

and muons measured on the ground:
\[
E_\mu=N_\mu(E>1\:GeV)\overline{E_\mu},
\]
where the average energy of muons $\overline{E_\mu} $ is taken from the MSU array data \cite{Khrenov}.

Residuary energy fractions transferred to neutrinos $E_\nu$, nucleons $E_h$ etc., unmeasurable with this array, are estimated using model calculations. The resulting apportioning of the primary energy 10$^{18}$ eV is given in Table~\ref{table:Balance}.
\begin{table}
\caption{The primary energy portions gone with EAS components. $E_0=10^{18}$ eV. $\theta=0^0$. Notations: $E_i$ is ionization loss of electrons in the atmosphere;
$E_g$ is ionization loss of electrons in the ground;
$E_{\mu+\nu}$ is energy transferred to muons and neutrinos;
$E_h$ is energy carried by the nuclear active component.
$E_0=E_i+E_g+E_{\mu+\nu}+E_h.$}
\begin{center}
\begin{tabular}{|l|c|c|}
\hline
Energy deposit & The portion   & Experimental\\
channel        & of energy, \% & uncertainty, \%\\\hline
$E_i$ & 78 & 30\\
$E_g$ &  9 & 30\\
$E_{\mu+\nu}$ &  9 & 16\\
$E_h$  &  4 & 20\\
\hline
\end{tabular}
\end{center}
\label{table:Balance}
\end{table}

\begin{figure}[htb]
\includegraphics[width=\columnwidth]{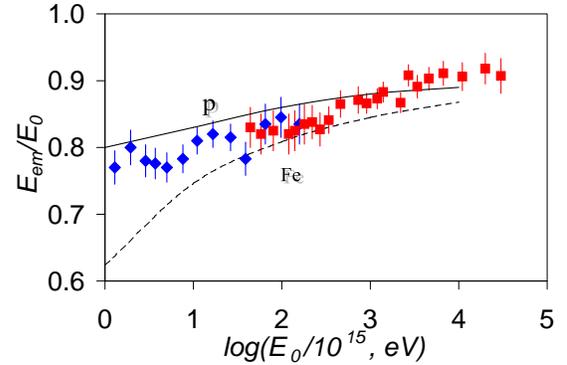}
\caption{The electromagnetic component energy estimation from the Yakutsk
array data: autonomous sub-array (rhombuses), the main array
(squares); P, Fe curves are $\delta$-model results.}
\label{fig:Eem}
\end{figure}

The energy fraction carried by electromagnetic component appears to be the basic contribution to the total energy of the shower, and its energy dependence (measured with Cherenkov light detectors + scintillators of the Yakutsk array) is illustrated in Fig.~\ref{fig:Eem}, together with $\delta$-model estimation under superposition assumption for the primary proton and iron nucleus.

Because the air Cherenkov light total flux, and electron and muon
number of the shower are experimental values measured on the
ground, only about 10\% of the primary energy $E_0=10^{18}$ eV is
calculated using the model assumptions. So we consider the energy
estimation algorithm in use in the Yakutsk group to be
model-independent in the first approximation.

\begin{figure}[htb]
\includegraphics[width=\columnwidth]{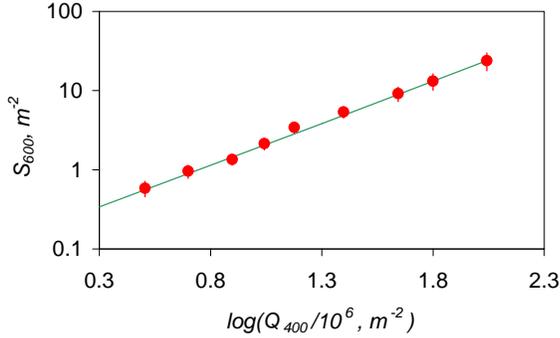}
\caption{Linear correlation between charged particle density at 600 m from the core ($S_{600}$) and air Cherenkov light intensity at 400 m
($Q_{400}$) measured in the same showers. $\bar{\theta}=10^0$.}
\label{fig:Correl}
\end{figure}

Moonless nights when air Cherenkov light measurements are possible
constitute $\sim10\%$ of the observation period. In order to evaluate the
primary energy of the bulk of showers, the linear correlation
\[
\log_{10}S_{600}=1.05\log_{10}Q_{400}-0.79;\: R^2=0.997
\]
is used between the charged particle density at 600 m from the shower
core and the light intensity at 400 m from the core,
$Q_{400}$ (Fig.~\ref{fig:Correl}) which, in turn, is related to the total flux of the Cherenkov light on the ground \cite{Mono}.

Experimental uncertainties in EAS component energies estimated using the Yakutsk array data are summarized in Table~\ref{table:Balance}. The main contribution arise from $\delta E_i$ which is formed by uncertainties in atmospheric transparency (15\%), detector calibration (21\%) and the total light flux measurement
(15\%). Errors in estimation of $N_e, \lambda_e, N_\mu+N_\nu$
determine the next two items (for ionization in the ground and
$\delta E_{\mu+\nu}$). Resultant energy estimation uncertainty is
the sum of all errors weighed with the second column of the Table~\ref{table:Balance}: $\delta E_0 \sim 30\%$. Extra 20\% is added due to $S_{600}-Q_{400}$ conversion uncertainty.

The conversion relation from $S_{300}/S_{600}$ to primary energy was re-examined in 2003 \cite{Prav03}. For vertical showers ($x=1020\: g/cm^2$) we have:
\begin{eqnarray}
E_0=(6.5\pm 1.6)\times 10^{16}S_{300}(0^0)^{0.94\pm0.02},\nonumber\\
E_0=(4.6\pm 1.2)\times 10^{17}S_{600}(0^0)^{0.98\pm0.02}.
\label{Eq:ERho}
\end{eqnarray}
Uncertainty of the conversion factor is mainly due to absolute calibration error of the Cherenkov light detector and optical transparency estimation error of the atmosphere. These errors affect the intensity of CR flux but not the shape of the spectrum.

\begin{figure}[htb]
\includegraphics[width=\columnwidth]{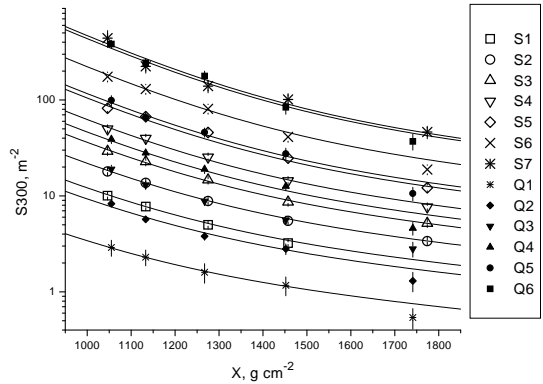}
\caption{$S_{300}$ as a function of $x=1020/\cos\theta$ for different CR intensities. Open symbols ($S_i$) are equi-intensity method results, filled ones ($Q_i$) are derived fixing $Q_{400}$.}
\label{fig:EquiCuts}
\end{figure}

The observed densities $S_{300}/S_{600}$ at zenith angle, $\theta$, are connected to the vertical direction values along attenuation curve \cite{Iv01}. In order to measure the attenuation length of these densities for fixed primary energy, we have used two different methods - well-known equi-intensity cut method, and fixing the Cherenkov light intensity at 400 m from the core as the equivalent of the primary energy, taking into account the light absorption in the atmosphere. In Fig.~\ref{fig:EquiCuts} the results are given. Experimental points are consistent with each other for the two methods used and can be described by the sum of two components - soft (electrons, attenuation length $\lambda_e=200\: g/cm^2$) and hard (muons, $\lambda_\mu=1000\: g/cm^2$):
\begin{eqnarray}
S_{300}(\theta)=S_{300}(0^0)((1-\beta_{300})exp((x_0-x)/\lambda_e)+\nonumber \\
+\beta_{300}exp((x_0-x)/\lambda_\mu)),
\end{eqnarray}
where $\beta_{300}$ is the hard component fraction. Attenuation curve for $S_{600}$ is the same but $\beta$ is different:
\begin{eqnarray}
\beta_{300}=(0.563\pm 0.032)S_{300}(0^0)^{-0.185\pm 0.02},\nonumber\\
\beta_{600}=(0.62\pm 0.006)S_{600}(0^0)^{-0.076\pm 0.03}.
\end{eqnarray}

Previously we used to apply another attenuation curve in order to estimate the primary energy \cite{Mono} which leads to divergence in energies of inclined showers rising with zenith angle.

Four showers in the energy range $E_0\geq10^{20}$ eV are given in Table~\ref{table:2} (one event is added slightly below the threshold because the energy estimation error is larger than tiny difference).

\begin{table}[htb]
\caption{Largest EAS events detected with the Yakutsk array}
\label{table:2}
\begin{tabular}{@{}cccccc}
\hline
  Date & $\theta^0$&$\log E_0$&$\delta E_0$,\%& $b^0$& $l^0$ \\
\hline
18.02.04 &   47.7   &  20.16  &      42      & 16.3 & 140.2 \\
07.05.89 &   58.7   &  20.14  &      46      &  2.7 & 161.6 \\
21.12.77 &   46.0   &  20.01  &      40      & 50.0 & 220.6 \\
15.02.78 &    9.6   &  19.99  &      32      & 15.5 & 102.0 \\
\hline
\end{tabular}\\[2pt]
\end{table}

\section{The energy spectrum measured with the Yakutsk array}
\begin{figure}[htb]
\includegraphics[width=\columnwidth]{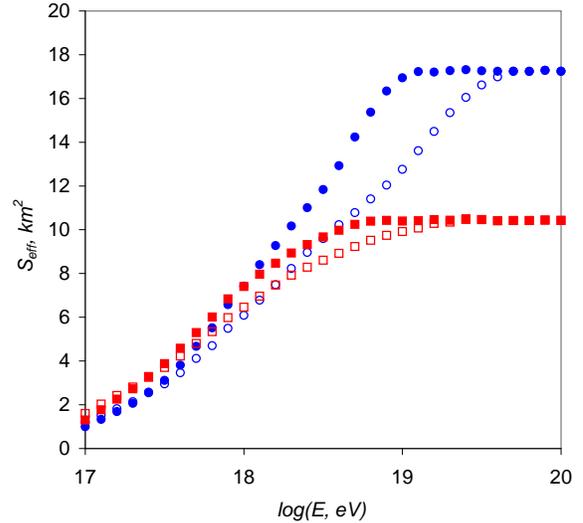}
\caption{Array area for vertical showers vs $E_0$. Circles are for the period before 1990; rectangles are for after the array was contracted. Two approximation functions of lateral distribution of charged particles are used: formula (\ref{Eq:Greisen}) (filled signs) and formula (\ref{Eq:Afan}) (open signs).}
\label{fig:SefFix}
\end{figure}

\begin{figure}[htb]
\includegraphics[width=\columnwidth]{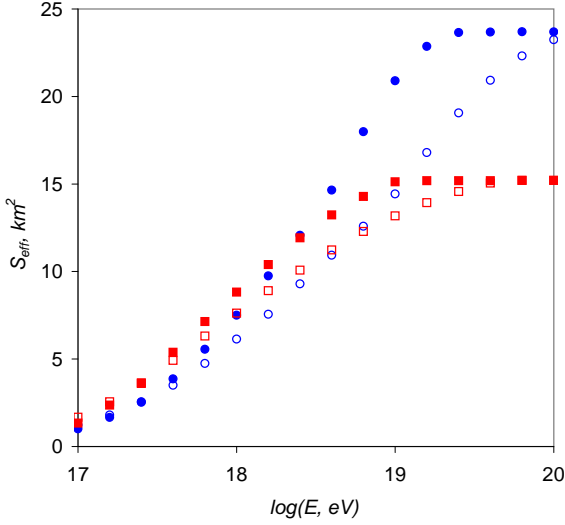}
\caption{The same areas but for extended perimeter. Notations are the same.}
\label{fig:SefExt}
\end{figure}

To derive $S_{eff}T\Omega$ - the product of array acceptance area, observation periods duration and solid angle, we have used $\Omega=\pi$; the real duration of the array duty time periods as $T$; and have calculated the area $S_{eff}(S_{300/600},\theta)$ limited within the array perimeter. The array configuration has been changed several times since 1974; the actual perimeter was used in particular time period. For the highest-energy showers $E_0>E_{GZK}$ the extended perimeter was applied. In this case the array area is increased $\sim1.4$ times while the shower parameters reconstruction is still possible with acceptable error. Energy dependence of the array area calculated for the vertical showers in the case of strict perimeter and extended one is shown in Fig.~\ref{fig:SefFix} and Fig.~\ref{fig:SefExt} correspondingly. The same procedure is applicable at each zenith angle for inclined EAS because $S(\theta)\neq S(0^0)\sec\theta$.


\begin{figure}[htb]
\includegraphics[width=\columnwidth]{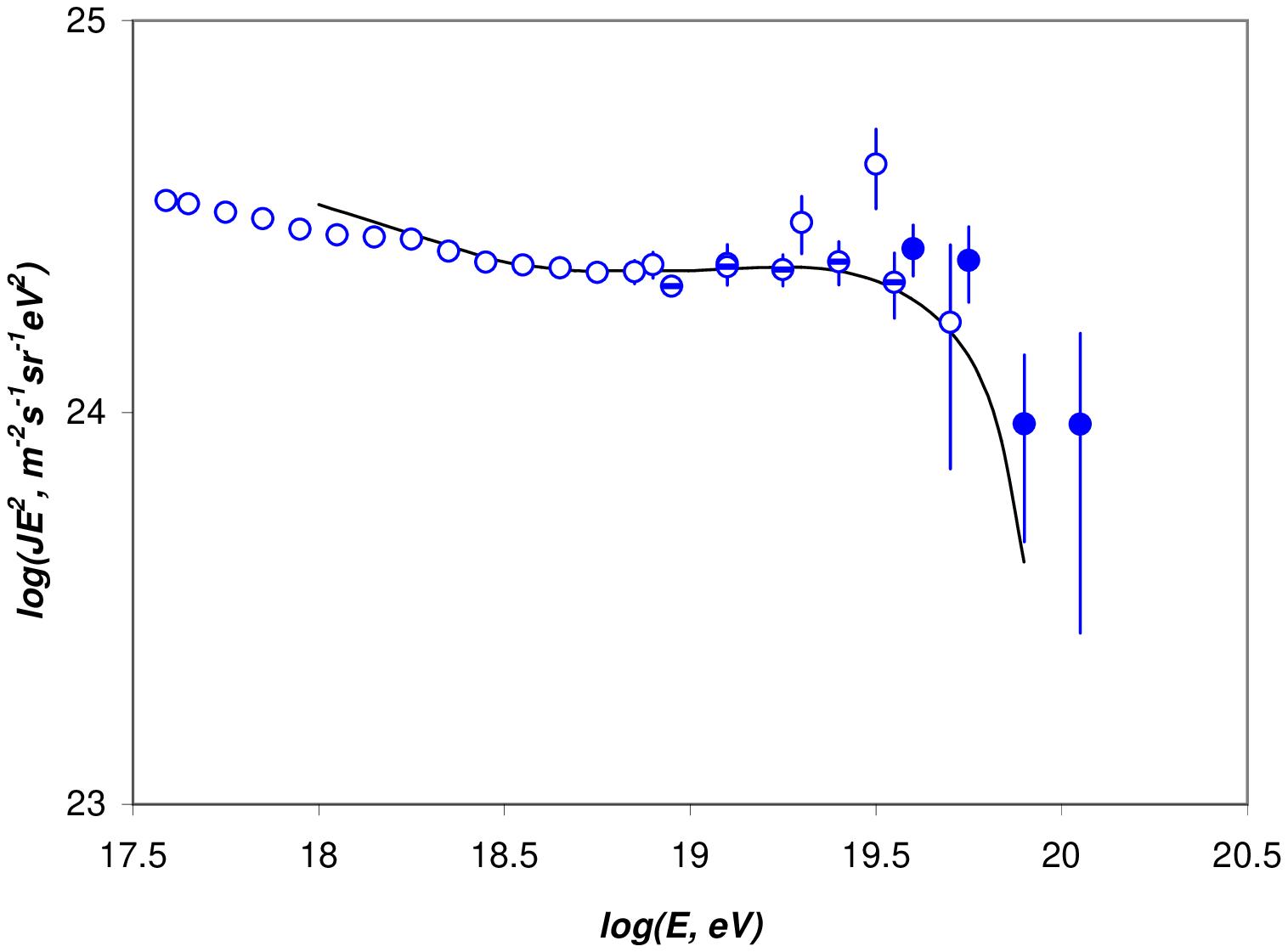}
\caption{Integral energy spectrum of CR detected with the Yakutsk array.
Data selected within array perimeter: open circles (trigger-500); ominus (trigger-1000);
filled circles (trigger-1000, extended perimeter).}
\label{fig:IntSpectr}
\end{figure}

In Fig.~\ref{fig:IntSpectr} resulting integral energy spectrum of cosmic rays is given composed of showers above $3\times10^{17}$ eV, with zenith angles $\theta<60^0$. We confirm with the present dataset analysis the existence of an 'ankle' feature in the shape of spectrum around $10^{19}$ eV, i.e. the flattening in the spectral index revealed earlier by all the groups working in this energy range \cite{NW}. At the highest energy end of the spectrum we still have a deficit in agreement with the expected flux cutoff due to interactions of CRs with relic microwave background. As an example of the model calculation results, the curve in Fig.~\ref{fig:IntSpectr} shows UHE proton spectrum \cite{Berez} in the model with uniform distribution of extragalactic sources with $E_{1/2}=6.2\times10^{19}$ eV.

\begin{figure}[htb]
\includegraphics[angle=90,width=\columnwidth]{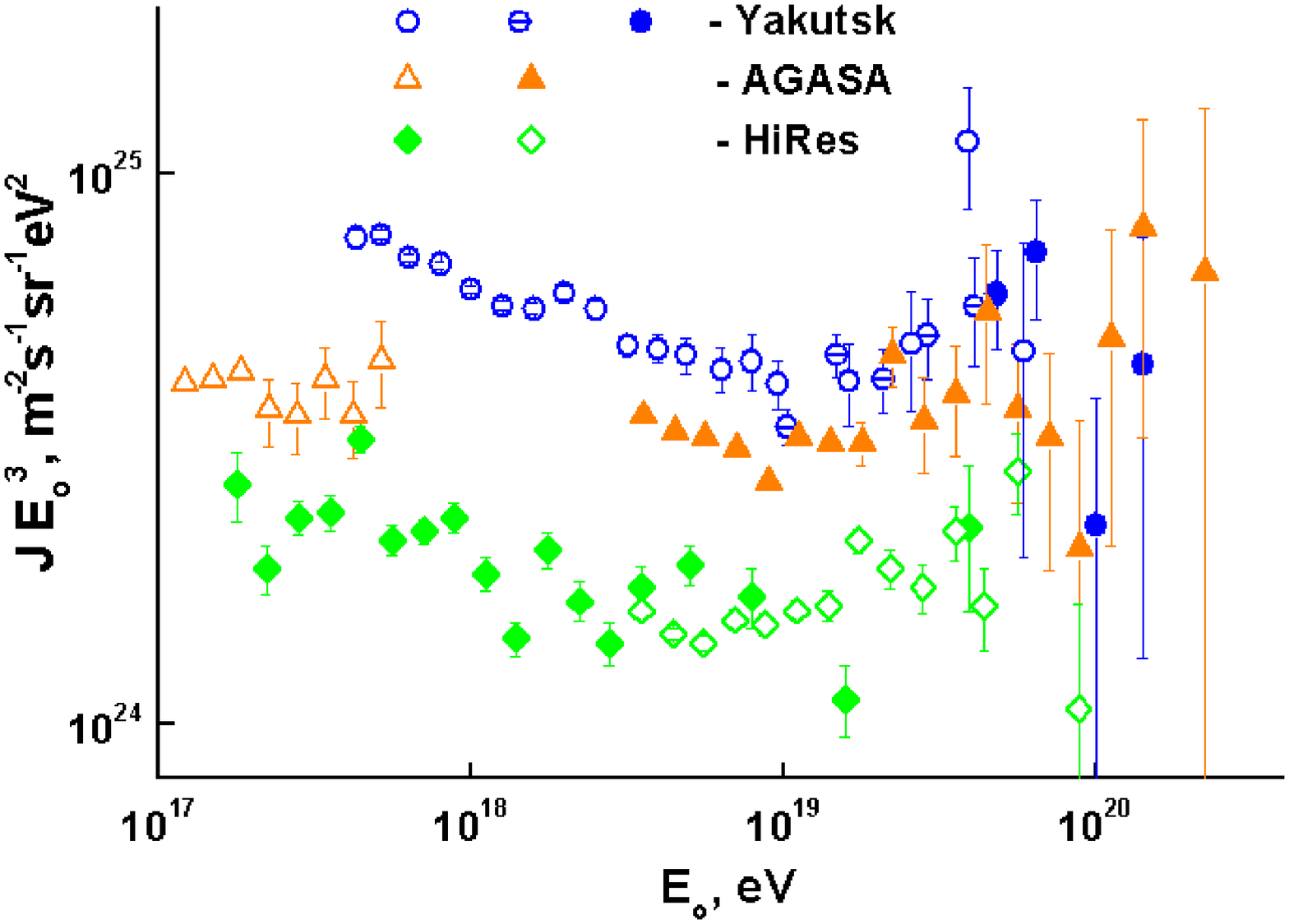}
\caption{Differential spectrum of UHECR, multiplied by $E_0^3$. For the Yakutsk array data notations are the same as in Fig.~\ref{fig:IntSpectr}. Open triangles show the Akeno \cite{Akeno} result, filled ones - AGASA \cite{AGASA}. Open rhombuses are for HiRes-I data and filled ones for HiRes-II \cite{HiRes}.}
\label{fig:DifSpectra}
\end{figure}

Comparison with other spectra measured with AGASA and HiRes arrays is shown in Fig.~\ref{fig:DifSpectra}. Differential intensities in energy bins measured in Yakutsk and statistical errors of the data are given in  Tables~\ref{table:T500},~\ref{table:T1000},~\ref{table:T1000X} for usability of the spectrum. In the energy range below $10^{18}$ intensities are highest of the Yakutsk array data. Most likely it is due to different energy estimation technique used in three cases - resultant energies diverge. In AGASA case the model dependent relation between $S_{600}$ and $E_0$ is used, HiRes team measures the fluorescence light flux which is connected to ionization loss of electrons in atmosphere. Our relations (\ref{Eq:ERho}) for vertical showers give energy higher than model calculations by (30-40)\% at $E_0\sim5\times10^{17}$ eV and by (15-20)\% at $E_0\sim5\times10^{19}$ eV.

\begin{table}
\caption{Cosmic ray flux measured under trigger-500, within array perimeter.}
\begin{tabular}{rrr}\hline
$E_0$ & $J(E)\times E^3$ & $\delta J$\\ \hline
5.89E+19 & 4.76E+24 & 2.75E+24\\
3.89E+19 & 1.15E+25 & 2.86E+24\\
2.57E+19 & 4.90E+24 & 1.19E+24\\
1.62E+19 & 4.20E+24 & 7.09E+23\\
9.55E+18 & 4.15E+24 & 4.15E+23\\
7.94E+18 & 4.56E+24 & 5.29E+23\\
6.31E+18 & 4.41E+24 & 4.14E+23\\
4.90E+18 & 4.69E+24 & 3.42E+23\\
3.98E+18 & 4.81E+24 & 2.79E+23\\
3.16E+18 & 4.89E+24 & 2.30E+23\\
2.51E+18 & 5.68E+24 & 1.99E+23\\
2.00E+18 & 6.08E+24 & 1.70E+23\\
1.58E+18 & 5.68E+24 & 1.36E+23\\
1.26E+18 & 5.75E+24 & 1.09E+23\\
1.00E+18 & 6.17E+24 & 1.05E+23\\
7.94E+17 & 6.84E+24 & 1.03E+23\\
6.31E+17 & 7.05E+24 & 1.06E+23\\
5.13E+17 & 7.74E+24 & 1.39E+23\\
4.27E+17 & 7.64E+24 & 2.75E+23\\ \hline
\end{tabular}
\label{table:T500}
\end{table}
\begin{table}
\caption{Cosmic ray flux measured under trigger-1000, within array perimeter.}
\begin{tabular}{rrr}\hline
$E_0$ & $J(E)\times E^3$ & $\delta J$\\ \hline
4.07E+19 & 5.75E+24 & 1.29E+24\\
2.88E+19 & 5.08E+24 & 8.59E+23\\
2.09E+19 & 4.23E+24 & 5.58E+23\\
1.48E+19 & 4.70E+24 & 4.23E+23\\
1.02E+19 & 3.47E+24 & 2.53E+23\\\hline
\end{tabular}
\label{table:T1000}
\end{table}
\begin{table}
\caption{Cosmic ray flux measured under trigger-1000, within extended array perimeter.}
\begin{tabular}{rrr}\hline
$E_0$ & $J(E)\times E^3$ & $\delta J$\\ \hline
1.41E+20 & 4.50E+24 & 3.18E+24\\
1.00E+20 & 2.30E+24 & 1.628E+24\\
6.43E+19 & 7.19E+24 & 1.75E+24\\
4.81E+19 & 6.04E+24 & 1.23E+24\\\hline
\end{tabular}
\label{table:T1000X}
\end{table}

At the highest energies our intensities became closer to AGASA data, but GZK cutoff feature is different. The slope of the spectrum above $10^{20}$ eV seems rising in the case of the Yakutsk array data as well as for HiRes spectrum; on the contrary, AGASA data exhibit no cutoff. But the probability to detect three or less events above $10^{20}$ eV in Yakutsk is 15\%, while 6 events are expected according to AGASA's intensity. Therefore, it seems that one has to wait until Auger/TA data in order to decide whether there is the GZK cutoff or not.

\begin{figure}[htb]
\includegraphics[width=\columnwidth]{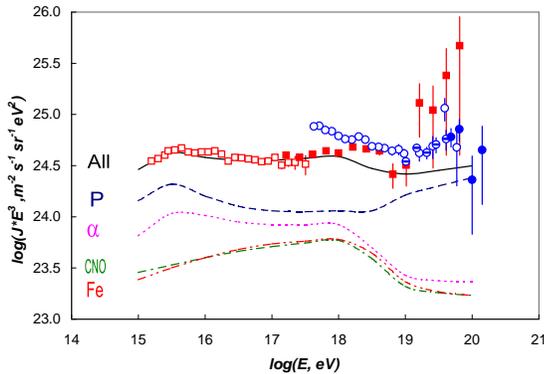}
\caption{Two spectra measured in Yakutsk using different experimental technique. Scintillator detector data are shown by circles as in previous figures; squares are the air Cherenkov light measurement results \cite{CERN}. Curves are anomalous diffusion model results \cite{Lagutin}.}
\label{fig:SpectrCher}
\end{figure}

In addition to the charged particle detection on the ground, we have another technique at the Yakutsk array - the air Cherenkov light measurement, which can be used to draw out the CR spectrum in independent way ~\cite{CERN}. These two spectra are shown in Fig.~\ref{fig:SpectrCher} covering the wide energy region from $10^{15}$ to $10^{20}$ eV. The Cherenkov detector sub-array is covering only about $3$ km$^2$ area, so it is convenient at energies $E<E_{GZK}$, while scintillator stations can measure the spectrum above this limit. At the low energy threshold of the main array $\sim10^{18}$ eV we have a discrepancy in intensity measured using two methods. This may be caused by $S_{300}$ uncertainty increasing in the vicinity of the threshold.

Two features of the measured spectrum - a 'knee' (at $E\sim3\times10^{15}$ eV) and 'dip' around $E\sim10^{19}$ eV can be explained in the model of anomalous diffusion of galactic cosmic rays in fractal interstellar medium ~\cite{Lagutin} . In this model the mass composition is predicted in the energy range $10^{14}-10^{20}$ eV; EAS maximum depth vs. energy is simulated by CORSIKA v.6.0 code using QGSjet model assumptions. Resulting spectra of the primary nuclei groups are shown by curves in Fig.~\ref{fig:SpectrCher}, together with the total spectrum marked 'All'.

\section{Conclusions}
We have measured the energy spectrum of cosmic rays below and surrounding the GZK cutoff region. Air Cherenkov light measurements provide us the model-independent approach to the primary energy estimation using the total light flux connected to the ionization loss of electrons in atmosphere. Other appreciable energy fractions are evaluated on the basis of the total number measurement of electrons and muons on the ground. An updated function to approximate the lateral distribution of charged particles leads to increased average density $S_{600}$. As a result, we have three showers detected with energies above $10^{20}$ eV including a new event detected 18.02.04. Two energy spectra measured with the Yakutsk array cover the range $10^{15}$ to $\sim10^{20}$ eV and exhibit the 'knee' and 'dip' features in agreement with other arrays. The spectrum tail above $E_{GZK}$ is consistent with the GZK cutoff contrary to AGASA result, although the statistical significance of this conclusion is insufficient due to a few showers detected above $10^{20}$ eV.

\section*{Acknowledgments}
We are grateful to the Yakutsk array staff for the essential contribution to data acquisition and analysis. AAI would like to thank CRIS 2004 organizers for invitation and hospitality. This work is supported by RFBR grants \#02-02-16380, \#03-07-90065, \#03-02-17160 and INTAS grant \#03-51-5112. Observations at the Yakutsk array is supported by Russian Ministry of Sciences and Education under the program \#01-30.


\begin{thebibliography}{9}
\bibitem{G} K. Greisen, Phys. Rev. Lett. 16 (1966) 748.
\bibitem{ZK} G.T. Zatsepin and V.A. Kuzmin, JETP Lett. 4 (1966) 144.
\bibitem{Cat1988} N.N. Efimov et al., Catalogue of HECR N3, World Data Center C2, Japan (1988) 56.
\bibitem{Mono} M.N. Dyakonov  et al., Cosmic Rays of Extremely High Energy (Nauka, Novosibirsk) (1991).
\bibitem{Tamm} I.E. Tamm, Journ. of Phys. (USSR) 1 (1939) 439.
\bibitem{HiResCurve} T. Abu-Zayyad et al., ApPh 16 (2001) 1.
\bibitem{Erlykin} A.D. Erlykin and A.W. Wolfendale, astro-ph/0112553.
\bibitem{Song} C. Song et al., ApPh 14 (2000) 7.
\bibitem{CERN} A.A. Ivanov, S.P. Knurenko and I.Ye. Sleptsov, Nucl.
Phys. B (Proc. Suppl.) 122 (2003) 226.
\bibitem{Afan} B.N. Afanasiev et al., in Proc. Worksh. Techn. Study EHECR, Tokyo (1993) 35.
\bibitem{Prav03} A.V. Glushkov et al., Proc. 28th ICRC, Tsukuba 1 (2003) 393.
\bibitem{Iv01} A.A. Ivanov et al., Izv. RAN, ser. fiz. 65 (2001) 1221.
\bibitem{NW} M. Nagano and A.A. Watson, Rev. Mod. Phys. 72 (2000) 689.
\bibitem{Akeno} N. Hayashida et al., Proc. Int. Symp. on EHECR: Astroph. Future Observ., Tokyo (1996) 17.
\bibitem{AGASA} N. Sakaki et al., Proc. 27th ICRC, Hamburg 1 (2001) 333.
\bibitem{Khrenov} B.A. Khrenov, Muons in EAS and nuclear interactions of CR particles in air (in Russian). Doct. Phys. Math. Thesis, MSU (1986).
\bibitem{HiRes} T. Abu-Zayyad et al., astro-ph/0208301.
\bibitem{Berez} V. Berezinsky, A. Gazizov and S. Grigorieva, astro-ph/0302483.
\bibitem{Lagutin} A.A. Lagutin et.al., Nucl. Phys. B (Proc. Suppl.) 97 (2001) 267.
\end{thebibliography}
\end{document}